# $^{129}$I and $^{247}$Cm in Meteorites Constrain the Last Astrophysical Source of Solar *r*-process Elements

Benoit Côté[1,2,3*], Marius Eichler[4], Andrés Yagüe[1], Nicole Vassh[5], Matthew R. Mumpower[6,7], Blanka Világos[1,2], Benjámin Soós[1,2], Almudena Arcones[4,8], Trevor M. Sprouse[5,6], Rebecca Surman[5], Marco Pignatari[9,1], Mária K. Pető[1], Benjamin Wehmeyer[1,10], Thomas Rauscher[10,11], Maria Lugaro[1,2,12]

[1]Research Centre for Astronomy and Earth Sciences, Eötvös Loránd Research Network, Konkoly Observatory, 1121 Budapest, Hungary

[2]Institute of Physics, Eötvös Loránd University, 1117 Budapest, Hungary

[3]National Superconducting Cyclotron Laboratory, Michigan State University, East Lansing 48824, USA

[4]Institut für Kernphysik, Technische Universität Darmstadt, 64289 Darmstadt, Germany

[5]Department of Physics, University of Notre Dame, Notre Dame 46556, USA

[6]Theoretical Division, Los Alamos National Laboratory, Los Alamos 87545, USA

[7]Center for Theoretical Astrophysics, Los Alamos National Laboratory, Los Alamos 87545, USA

[8]GSI Helmholtzzentrum fur Schwerionenforschung GmbH, 64291 Darmstadt, Germany

[9]E.A. Milne Centre for Astrophysics, University of Hull, Hull HU6 7RX, United Kingdom

[10]Centre for Astrophysics Research, University of Hertfordshire, Hatfield AL10 9AB, United Kingdom

[11]Department of Physics, University of Basel, 4056 Basel, Switzerland

[12]Monash Centre for Astrophysics, School of Physics and Astronomy, Monash University, VIC 3800, Australia

*Correspondence to: benoit.cote@csfk.org

**Abstract: The composition of the early Solar System can be inferred from meteorites. Many elements heavier than iron were formed by the rapid neutron-capture process (*r* process), but the astrophysical sources where this occurred remain poorly understood. We demonstrate that the near-identical half-lives (≃ 15.6 Myr) of the radioactive *r*-process nuclei $^{129}$I and $^{247}$Cm preserve their ratio, irrespective of the time between production and incorporation into the Solar System. We constrain the last *r*-process source by comparing the measured meteoritic $^{129}$I / $^{247}$Cm = 438 ± 184 to nucleosynthesis calculations based on neutron star merger and magneto-rotational supernova simulations. Moderately neutron-rich conditions, often found in merger disk ejecta simulations, are most consistent with the meteoritic value. Uncertain nuclear physics data limit our confidence in this conclusion.**

The rapid neutron-capture process (*r* process) is the source of half the naturally occurring elements heavier than iron (*1*), including iodine, europium, gold, platinum, and the actinides. However, the astrophysical sites where *r*-process elements were synthesized and the physical conditions at these sites are not well constrained.

The gravitational wave event GW170817 (*2*), the identification of its electromagnetic counterpart, and the inference of lanthanide elements in the ejecta (*3*), showed that neutron star mergers can synthesize at least some *r*-process elements. GW170817 provided only limited information on the nucleosynthesis process, as only one specific element (strontium) has been identified in its spectrum (*4*). More detailed isotopic information for *r*-process nucleosynthesis is recorded in the composition of the Solar System. Analysis of primitive meteorites has produced abundance determinations for all stable isotopes (*5*), whereas abundances derived from stellar spectra typically provide elemental abundances only.

The Solar System's stable isotopes include contributions from multiple nucleosynthetic events (supernovae, compact binary mergers, etc.) that occurred at any time between the birth of the Milky Way and the formation of the Sun. This evolution is difficult to model, but can be simplified by considering radioactive isotopes with half-lives of several million years (Myr). Analysis of meteorites shows that such isotopes were present at the formation time of the first solids (the calcium-aluminium-rich inclusions, CAIs) in the early Solar System (*6*). Because those radioactive isotopes have all decayed over the lifetime of the Solar System, their initial abundances are inferred from excesses of the daughter isotopes they decay into. Radioactive isotopes reflect a smaller number of nucleosynthesis events than stable isotopes, specifically the events which occurred shortly prior to the formation of the Sun. We consider the early Solar System abundances of two radioactive isotopes with half-lives of 15.7 and 15.6 Myr, respectively: $^{129}$I and the heavier actinide isotope, $^{247}$Cm. We adopt abundances of these isotopes (Table 1) from previously published analyses of meteorites (*7–9*), where they are reported as ratios with reference isotopes: $^{129}$I / $^{127}$I and $^{247}$Cm / $^{235}$U.

Comparing these isotopic ratios directly with predictions from simulations and determining the nucleosynthetic sources that enrich interstellar gas with heavy elements is highly uncertain. The abundance ratio $^{129}$I / $^{127}$I has a stable isotope in the denominator, the abundance of which depends on the complete galactic enrichment history prior to the formation of the Solar System. This ratio is therefore affected by uncertainties on the star formation history, the amount of interstellar gas within the Milky Way, and the amount of $^{127}$I removed from the interstellar gas by galactic outflows (*14*). The $^{247}$Cm / $^{235}$U ratio is less affected by those uncertainties because $^{235}$U has a half-life of 704 Myr, which is short relative to the ~ 8 − 9 billion years of galactic enrichment prior to the formation of the Sun. The $^{247}$Cm / $^{235}$U ratio is still affected by the uncertain time interval between their synthesis and incorporation into the early Solar System. This delay is ~ 100 − 200 Myr for *r*-process isotopes (*13*), during which $^{247}$Cm and $^{235}$U decay exponentially. Because their half-lives differ by a factor of 50, the $^{247}$Cm / $^{235}$U abundance ratio diverges from its original value before being locked into the Solar System.

Enrichment of the interstellar gas from which the Solar System formed was not continuous but stochastic (*15*). It is therefore unknown how many enrichment events are recorded in the isotopic ratios derived from meteorites. Because the radioactive abundances from each event decayed for an unknown amount of time, the relative contributions are even more uncertain.

Using the $^{129}$I / $^{247}$Cm abundance ratio bypasses those uncertainties because of the combination of two properties. First, $^{129}$I and $^{247}$Cm have the same half-life, within uncertainties, so their ratio is not strongly affected by decay over time. Second, both isotopes are short-lived compared to the average time elapsed between *r*-process events, so their ratio probably reflects only one event (Supplementary Text). Fig. 1 shows a simulation of how these isotope ratios vary over time. $^{129}$I / $^{247}$Cm always stays close to its production ratio, while $^{129}$I / $^{127}$I and $^{247}$Cm / $^{235}$U vary by orders of magnitude. Different astrophysical sources could have synthesized a range of $^{129}$I / $^{247}$Cm abundance ratios throughout the history of the Galaxy, but only one event is likely recorded in meteorites for these isotopes. We determined the $^{129}$I / $^{247}$Cm ratio in the early Solar System (Table 1) using the reported $^{129}$I / $^{127}$I and $^{247}$Cm / $^{235}$U ratios, together with the $^{127}$I / $^{235}$U ratio of 189 (*5*). We find $^{129}$I / $^{247}$Cm = 438 ± 184, and interpret this value as reflecting the nucleosynthesis of the last *r*-process event that polluted the pre-solar nebula.

This value relies on our adoption of solar abundances commonly used in astronomy (*5*). Alternative measurements have reported an iodine abundance that is an order of magnitude lower (*17*), which would affect our conclusions. Adopting the lower value would make iodine less abundant than neighbouring isotopes. Our nucleosynthesis calculations (see below) do not predict this feature since they generally show smoother abundance trends between neighbouring species, which is more consistent with the higher abundance measurement (*5*). The meteoritic measurements (*17*) could be affected by heterogeneities on scales larger than the samples that were analyzed (the nugget effect) and by possible losses of noble gases produced from halogens such as iodine via the irradiation technique adopted for the measurements (Supplementary Text). We therefore prefer to adopt the higher value of the iodine abundance (*5*) (Supplementary Text).

We performed theoretical nucleosynthesis calculations to determine the $^{129}$I / $^{247}$Cm abundance ratios that would be produced in the physical conditions that occur in previous hydrodynamic simulations of potential *r*-process sites: neutron star – neutron star (NS-NS) mergers, neutron star – black hole (NS-BH) mergers, and core-collapse supernovae driven by strong magnetic fields and fast rotation [magneto-rotational supernovae, MR SN (*18*)]. In NS-NS and NS-BH mergers, matter is ejected in two ways: i) dynamical ejecta (*19,20*) that are driven by tidal forces and shocks that occur promptly during the merger, ii) disk ejecta (*21*) due to heating that unbinds matter from the disk that forms around the compact central remnant left after the merger, which is either a neutron star or a black hole. Table S1 lists details of the seven simulations we considered. Because *r*-process nucleosynthesis predictions are affected by large uncertainties from nuclear physics (*22–24*), we repeated our calculations using three different sets of nuclear reaction rates and three different models for the distribution of fission fragments (*13*). This generated nine nucleosynthetic model predictions that were applied to each of the seven hydrodynamic simulations, for a total of 63 calculations shown in Fig. 2.

In Fig. 2, we compare our predicted $^{129}$I / $^{247}$Cm ratios using different nuclear physics input with the meteoritic ratio. The uncertainties on the meteoritic ratio include both the uncertainty in the derivation of the early Solar System ratio (Table 1) and the uncertainty in the half-lives of $^{129}$I and $^{247}$Cm. We include the latter to account for the slight ratio variation that could have occurred during the time elapsed between the last *r*-process event and the condensation of the first solids in the early Solar System (*13*). Because $^{129}$I and $^{247}$Cm have substantially different atomic numbers, their relative abundances strongly depend on the physical conditions in which the *r*-process nucleosynthesis occurs. The predicted ratios shown in Fig. 2 vary by more than two orders of magnitude.

For the MR SN ejecta (*18*), the abundance ratio is always larger than 1000 because most of the ejecta are not sufficiently neutron-rich to produce enough actinides. Although other MR SN simulations may generate different results, models with alternative neutrino transport predict even lower production of actinides (*25*). MR SNe are expected to have occurred more often in the early Universe, due to higher stellar rotation (*26*), which makes MR SNe more likely to enrich very old stars than the Solar System. Collapsars are also a possible *r*-process site; these occur during the late evolution of some MR SNe when a black hole surrounded by an accretion disk forms. However their capacity to synthesize actinides (including $^{247}$Cm) is debated and ranges from substantial production (*27*) to no production (*28,29*).

For the NS-NS and NS-BH merger simulations, dynamical ejecta are dominated by very neutron-rich conditions, producing more actinides (such as $^{247}$Cm) relative to lighter nuclei ($^{129}$I in this case), compared to the other *r*-process scenarios (see also Fig. S1). As a result, the dynamical ejecta $^{129}$I / $^{247}$Cm ratios are all lower than 100, which is below the 2σ uncertainty of the meteoritic ratio. Merger simulations predict the presence of very neutron-rich material (*19,20*), however, the exact contribution of such conditions to the total ejecta is still unclear. Simulations of dynamical ejecta show a broad range of neutron richness (*30*).

The three NS-NS merger accretion-disk ejecta simulations give different results (Fig. 2). NS-NS disk 1 is consistent with the meteoritic value, NS-NS disk 2 partly overlaps with the 2σ uncertainty, while NS-NS disk 3 is below the 2σ uncertainty and therefore not compatible. Although these disk simulations represent a disk forming around an NS-NS remnant, NS-BH disk models can produce similar abundances (*21*).

We considered the combination of both dynamical and disk ejecta from a single binary merger (Supplementary Text). We found that the maximum contribution of dynamical ejecta is ~ 50 % (in mass fraction) in order to remain within the 2σ uncertainty of the meteoritic ratio. $^{129}$I and $^{247}$Cm in the early Solar System were likely synthesized by only one *r*-process event, but if two events contributed, the meteoritic ratio could be matched by a combination of dynamical ejecta with MR SN ejecta (see Fig. 2). However, such a mixture has an occurrence probability of less than 10% (Supplementary Text).

To test the sensitivity of these results to the input data, we performed 56 additional nucleosynthesis calculations on the dynamical ejecta (*19*) using a different nucleosynthesis code and a wider variety of input nuclear physics models (*13*). The vast majority of these models predict $^{129}$I / $^{247}$Cm ratios below 100 (Tables S2 and S3), which is consistent with the results presented in Fig. 2. In four of the 56 cases very neutron-rich dynamical ejecta reach the meteoritic ratio. The large range in predictions is due to the nuclear physics uncertainties. Our additional calculations support our conclusion that enrichment of the pre-solar nebula by very neutron-rich ejecta is often inconsistent with the meteoritic data. This result applies to the last *r*-process event that polluted the pre-solar nebula with radioactive isotopes, not to the collective contribution of all previous events that built up the stable *r*-process solar composition.

We have shown that the $^{129}$I / $^{247}$Cm abundance ratio can constrain the ejecta composition of the last *r*-process event that polluted the pre-solar nebula. This ratio is highly sensitive to the physical conditions in which $^{129}$I and $^{247}$Cm were synthesized. Our results suggest that moderately neutron-rich conditions are generally most consistent with meteoritic value. However, such conclusions are limited by solar abundance determinations and current uncertainties in the available hydrodynamical and nucleosynthesis models.

**Acknowledgments:** We thank Katharina Lodders, Uli Ott, and Jamie Gilmour for discussion. We thank the reviewers and the editorial team for improving the content of our manuscript. This work has benefited from discussions at the 2019 Frontiers in Nuclear Astrophysics Conference supported by the JINA Center for the Evolution of the Elements, and at conferences supported by the ChETEC (Chemical Elements as Tracers of the Evolution of the Cosmos) COST Action (CA16117, European Cooperation in Science and Technology). MP acknowledges access to VIPER, the University of Hull High Performance Computing Facility: **Funding:** BC, AY, BW, MKP, and ML were supported by the ERC Consolidator Grant (Hungary) funding scheme (Project RADIOSTAR, G.A. n. 724560). BC and ML were supported by the Hungarian Academy of Sciences via the Lendület project LP2014-17. BC, MRM, and MP acknowledge support from the National Science Foundation (NSF, USA) under grant No. PHY- 1430152 (JINA Center for the Evolution of the Elements). MRM was supported by the US Department of Energy through the Los Alamos National Laboratory and by the Laboratory Directed Research and Development program of Los Alamos National Laboratory under project number 20190021DR. Los Alamos National Laboratory is operated by Triad National Security, LLC, for the National Nuclear Security Administration of U.S. Department of Energy (Contract No. 89233218CNA000001). ME and AA acknowledge support from European Research Council through ERC Starting Grant No. 677912 EUROPIUM and Deutsche Forschungsgemeinschaft through SFB 1245. NV and RS were supported by the Fission In R-process Elements (FIRE) topical collaboration in nuclear theory, funded by the U.S. Department of Energy. AA was supported by the Helmholtz Forschungsakademie Hessen für FAIR. TS and RS were supported by the U.S. Department of Energy SciDAC collaboration TEAMS (DE- SC0018232). TS was supported by the Los Alamos National Laboratory Center for Space and Earth Science, which is funded by its Laboratory Directed Research and Development program under project number 20180475DR. MP acknowledges support to NuGrid from STFC through the University of Hull's Consolidated Grant ST/R000840/1; **Author contributions:** BC developed the concept, led and coordinated the collaboration, co-developed the statistical framework that led to Fig. 1, and participated in the writing and reviewing process. ME performed and analysed the nucleosynthesis calculations shown in Figs. 2 and S2, ran the sampling procedure shown in Fig. S1, and participated in the writing and reviewing process. AY performed the calculations shown in Table S6, helped develop the concept, co-developed the statistical framework that led to Fig. 1, and participated in the writing and reviewing process. NV performed and analysed the nucleosynthesis calculations shown in Tables S2 and S3, and participated in the writing and reviewing process. MRM participated in the reviewing process, and in the development of the PRISM code and the nuclear reaction rates used in PRISM. BV calculated the probability distributions shown in Figs. S4 and S5. BS ran the calculations shown in Fig. S3. AA, RS, MP and BW participated in the reviewing process. TMS participated in the development of the PRISM code. MKP participated in the interpretation of meteoritic abundances and in the writing reviewing process. TR participated in the development of the nuclear reaction rates used in WINNET, and in the reviewing process. ML calculated the early Solar System $^{129}$I / $^{247}$Cm ratio shown in Table 1, helped develop the concept, and participated in the writing and reviewing process; **Competing interests:** We declare no conflicts of interest; and **Data and materials availability:** The code used to calculate the isotopic ratios shown in Fig. 1 is available at https://github.com/AndresYague/Stochastic_RadioNuclides (*31*). The WINNET nucleosynthesis output and sampling code to reproduce Figs. 2, S1 and S2 are available on Zenodo (*32*). The code to reproduce Fig. S3 is available on Github at https://github.com/AndresYague/delta_stat_tau_code/tree/v1.0.0 (*33*). The Monte Carlo code used to calculate the distributions shown in Figs. S4 and S5 is available at https://github.com/AndresYague/IodineCurium_project_distributions (*34*). The PRISM nucleosynthesis output necessary to reproduce Tables S2 and S3 are available on Zenodo (*35*); this dataset is released under Los Alamos National Laboratory report number LA-UR-21-20444. The Monte Carlo code to generate Table S6 is available at https://github.com/AndresYague/IodineCurium_project_oneEvent (*36*). The trajectories of the dyn. ejecta (R) simulations were taken from https://compact-merger.astro.su.se/downloads_fluid_trajectories.html; the dyn. ejecta

(B) from (*20*); disk ejecta 1, 2, 3, from (*21*); and MR SN from (*18*). The WINNET code was developed by C. Winteler, F. Thielemann, O. Korobkin, M. Eichler, D. Martin, J. Bliss, M. Reichert, and A. Arcones; we do not have their permission to distribute it. The PRISM code is security restricted and unavailable for public release; contact M.R.M. for details.

**Supplementary Materials:**

Materials and Methods

Supplementary Text

Figures S1-S5

Tables S1-S6

References (*37-104*)

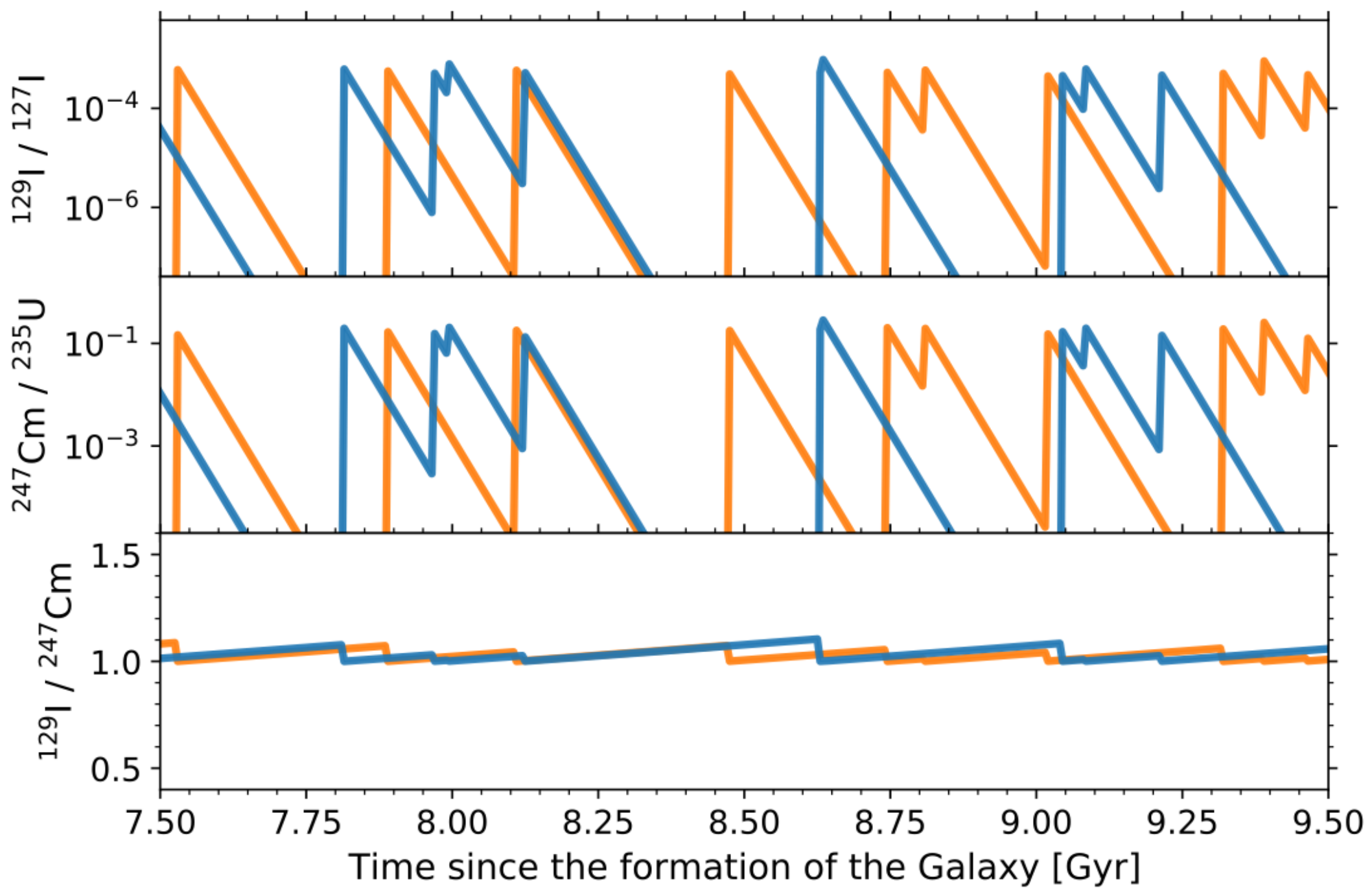

**Fig. 1. Simulated evolution of the abundance ratios $^{129}$I / $^{127}$I, $^{247}$Cm / $^{235}$U, and $^{129}$I / $^{247}$Cm in a parcel of Milky Way interstellar gas.** The time window shown encompasses the time when the Sun formed. Each peak is produced by an additional *r*-process event. The blue and orange lines show two arbitrary Monte Carlo realizations for the temporal distribution of those events (*16*). Each event is assumed to eject the same mass of $^{129}$I, $^{127}$I, $^{247}$Cm, and $^{235}$U, such that the production ratio is equal to one for all three isotopic ratios.

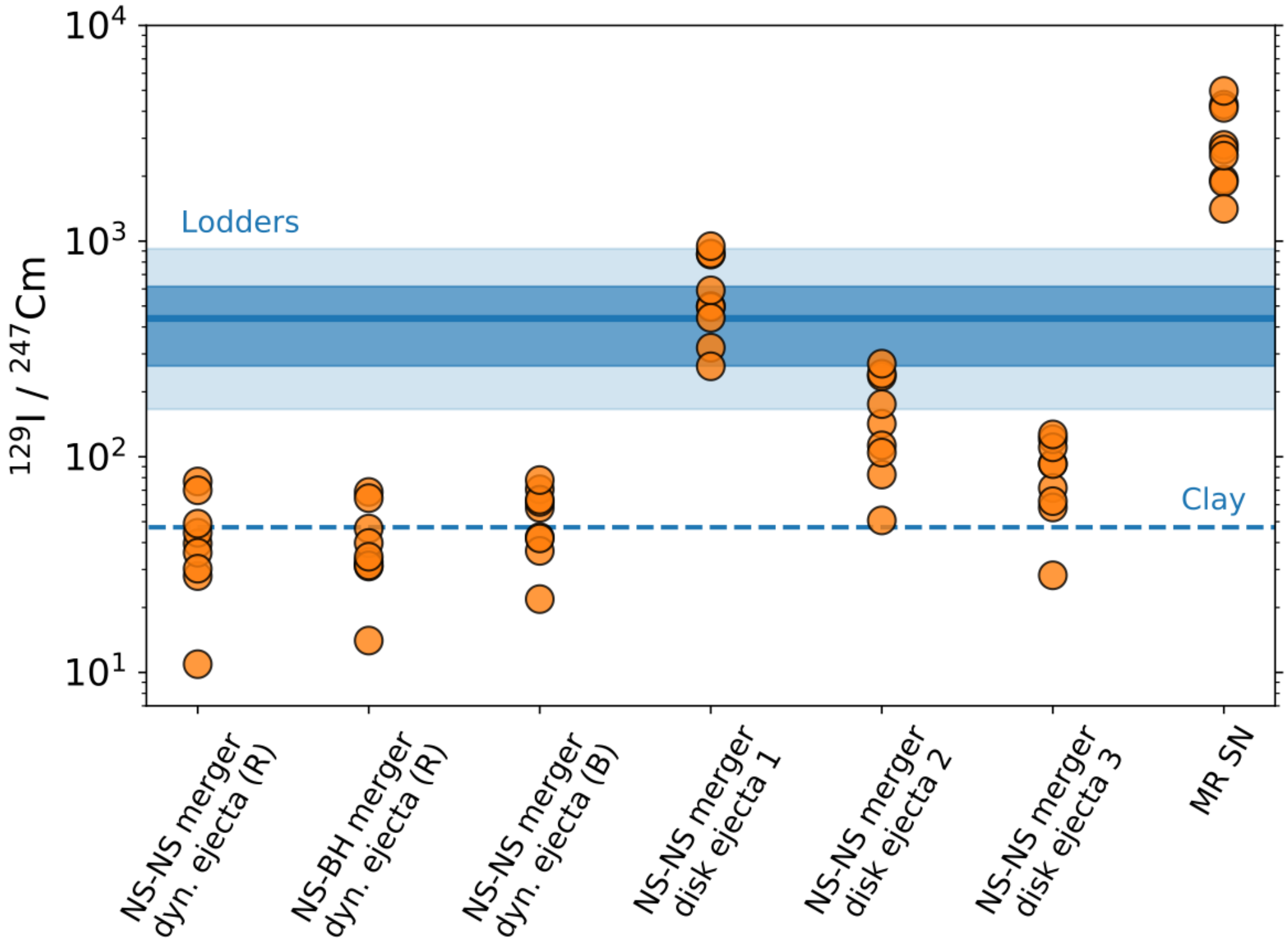

**Fig. 2. $^{129}$I / $^{247}$Cm abundance ratios predicted by theoretical *r*-process models.** The dots in each model denote different nuclear physics input (*13*). The blue solid horizontal line shows the meteoritic ratio and the shaded bands are its 1σ and 2σ uncertainty (*13*). The blue dashed horizontal line shows the meteoritic ratio when adopting alternative measurements (*17*).

**Table 1. Early Solar System isotopic ratios involving radioactive nuclei produced by the *r*-process.** Column 5 is the early Solar System ratio of each isotope listed in Column 1, relative to that in Column 3, with their half-lives (*10–12*) given in Columns 2 and 4, respectively. All uncertainties are 2σ (*13*).

| Short-lived radionuclide | Half-life (Myr) | Reference isotope | Half-life (Myr) | Early Solar System ratio | Refs. |
|---|---|---|---|---|---|
| $^{129}$I | 15.7 ± 0.8 | $^{127}$I | Stable | (1.28 ± 0.03) × $10^{-4}$ | (*7*) |
| $^{247}$Cm | 15.6 ± 1.0 | $^{235}$U | 704 ± 2 | (5.6 ± 0.3) × $10^{-5}$ | (*8, 9*) |
| $^{129}$I | 15.7 ± 0.8 | $^{247}$Cm | 15.6 ± 1.0 | 438 ± 184 | See text |

# Supplementary Materials for

## $^{129}I$ and $^{247}Cm$ in Meteorites Constrain the Last Astrophysical Source of Solar *r*-process Elements

Benoit Côté[1,2,3]*, Marius Eichler[4], Andrés Yagüe[1], Nicole Vassh[5], Matthew R. Mumpower[6,7], Blanka Világos[1,2], Benjámin Soós[1,2], Almudena Arcones[4,8], Trevor M. Sprouse[5,6], Rebecca Surman[5], Marco Pignatari[9,1], Mária K. Pető[1], Benjamin Wehmeyer[1,10], Thomas Rauscher[10,11], Maria Lugaro[1,2,12]

Correspondence to: benoit.cote@csfk.org

**This PDF file includes:**

Materials and Methods
Supplementary Text
Figs. S1 to S5
Tables S1 to S6

## Materials and Methods

Impact of Nuclear Physics Uncertainties on the Production of $^{129}$I and $^{247}$Cm

The *r*-process proceeds via synthesis of very neutron-rich isotopes that are currently inaccessible to experimental nuclear physics facilities. Calculations therefore must rely on theoretical data for the reaction and decay processes which determine the abundances of $^{129}$I and $^{247}$Cm. Because such theoretical data can vary greatly outside experimentally probed nuclei, this introduces uncertainties in the predicted $^{129}$I / $^{247}$Cm ratio.

In the case of $^{247}$Cm production, we find that this depends mostly on the model strength of the $N = 126$ shell closure (N refers to the number of neutrons), which determines how *r*-process material proceeds into the actinide region, along with the β-decay treatment which determines the decay path of nuclei which eventually populate this species. Our calculations using currently available theoretical fission treatments suggest fission does not greatly impact the production of $^{247}$Cm because it is less massive than isotopes for which fission influences *r*-process abundances. However α-decay feeding from heavier species introduces a small sensitivity to how fission is treated at higher mass numbers. In contrast, the production of $^{129}$I is influenced by the fission treatment via the fission fragment distributions of heavy, neutron-rich species which can contribute to the production of the second *r*-process peak (an enhancement in solar abundances at mass number ~130), but is also dependent on the predicted structure of the $N = 82$ shell closure. None of these nuclear data governing how $^{129}$I and $^{247}$Cm are populated in the *r* process have been determined experimentally.

Nucleosynthesis Models

The nucleosynthesis calculations presented in Fig. 2 were obtained with the nuclear network code WINNET (*18*). Fig. S2 shows the predicted ratios of the individual tracer particles that recorded the evolution of the physical conditions (density and temperature) as a function of time within different parts of the ejecta. Those temporal profiles are referred to as trajectories and represent the astrophysics inputs for our nucleosynthesis calculations. Fig. S2 also shows the sums of the trajectories. For each hydrodynamic simulation (see Table S1), we post-processed all tracer particles and included the additional heating caused by nuclear reactions, following previously-published methods (*37*). When available, we used experimental nuclear reaction rates, but most of the nuclear physics input necessary for *r*-process nucleosynthesis calculations is based on theoretical models that are uncertain for nuclei far from stability.

To test the impact of nuclear physics uncertainties, we repeated all of our calculations using nine combinations of nuclear physics input. For the first set, which we refer to as FRDM, we used the default JINA Reaclib reaction rate library (*38*) with reaction rates based on the Finite-Range Droplet Model (FRDM) (*39*). For the second set, FRDM(D3C*, which stands for adjusted model with Density-Dependent and Derivative Couplings), we used JINA Reaclib but replaced the default theoretical β-decay rates (*40,41*) with alternatives (*42*). For the third set, labeled DZ10, we used neutron-capture and charged-particle rates (and their reverse reactions) based on the Duflo-Zuker mass model with 10 parameters (*43*), together with β-decay rates (*40,41*). These three sets have been described previously [(*44*), their table 1]. Possible fission of actinides in very neutron-rich conditions (*45,46*) adds another layer of nuclear physics uncertainty to our predictions (*24*). To test this, for each input set (FRDM, FRDM(D3C*), and DZ10), we ran our nucleosynthesis calculations with three different fission fragment distribution models: Panov (*47*), K & T (*48*), and the ABLA07 model (*49*).

Fig. S2 shows the $^{129}$I / $^{247}$Cm ratios predicted using the nine nuclear physics model combinations described above. For each astrophysical source (i.e., hydrodynamic simulation), we show the isotopic composition of the total ejected mass and trajectories for individual parts of the ejecta that build up the total ejecta. Although the total ejecta are the most relevant for comparison with the meteoritic data, the different trajectories illustrate the wide range of physical conditions that can take place within a single *r*-process event.

Additional Calculations for Dynamical Ejecta

Figs. 2 and S2 show that very neutron-rich dynamical ejecta (NS-NS dyn. (R) and NS-BH dyn. (R)) produce $^{129}$I / $^{247}$Cm ratios below the meteoritic value. We performed additional calculations to investigate this conclusion using the nucleosynthesis network PRISM (*50*). We applied experimentally measured decay rates (*51*) where available. For theoretical nuclear data, we considered ten mass models (FRDM2012, FRDM1995, DZ33, TF, ETFSI, ETFSI-Q, HFB-17, HFB-21, SLy4, UNEDF0, see Table S4) matched with appropriate fission barrier sets when available [TF barriers (*60*) are applied with TF masses, ETFSI barriers (*61*) with ETFSI masses, HFB barriers (*62*) with HFB masses, and FRLDM barriers (*63*) in all other cases]. We used the statistical Hauser-Feshbach framework (*41,64,65*) to produce neutron-capture, β-decay, neutron-induced fission and β-delayed fission rates which are consistent with the theoretical masses and fission barriers (*24,66*). For every mass model we applied our default β-decay strength function (*67*) to determine our decay rates, but for a selection of mass models, we performed additional calculations to consider the alternative D3C* β-decay strength function (*42*). For the fission yields of heavy neutron-rich nuclei, we used GEF2016 (*68*).

The PRISM results for the $^{129}$I / $^{247}$Cm ratio with the different theoretical data sets are given in Tables S2 and S3. When using our default β-decay rates, the ranges given by the simulation trajectories are typically lower than the meteoritic $^{129}$I / $^{247}$Cm ratio, with ETFSI-Q predicting the highest possible value. The mass weighted average of all trajectories lies below the meteoritic ratio for all models. When using the alternative D3C* β-decay rates, however, the nuclear data combination of TF masses and fission barriers predicts $^{129}$I / $^{247}$Cm ratios that are consistent with the meteoritic value. Our calculations suggest that although very neutron-rich conditions in dynamical ejecta cannot be entirely ruled out as the last pre-solar *r*-process source, such conditions are often inconsistent with nucleosynthesis calculations.

Rarity of *r*-Process Events Near the Solar Neighborhood

A necessary input in modeling the amount of *r*-process isotopes introduced into the pre-solar nebula, and in determining whether the $^{129}$I / $^{247}$Cm ratio probes only one event, is the frequency of *r*-process events occurring in the solar neighborhood. For compact binary mergers (CBMs), there are constraints on their cosmic and galactic rates (*2,69,70*), but it remains uncertain how frequently a given parcel of gas is expected to be enriched by those mergers. Simple calculations suggest that regular core-collapse supernovae (CCSNe) could pollute a parcel of gas of the Milky Way every 5 − 10 Myr, when considering a "snowplough" scenario (*71*). Because CBMs are more rare than regular CCSNe by a factor of ~100 − 1000 (*72*), we expect the average time interval between CBM events to be at least 500 Myr for a given parcel of gas (*6*). If rare classes of CCSNe exist that are capable of producing heavy *r*-process elements up to Cm, their frequency is even more uncertain.

The evolution of $^{244}$Pu has been investigated (*15)* from the formation time of the Solar System until today, using a diffusive mixing prescription to spread radioactive isotopes throughout the interstellar medium. Their best fitting model to explain the variation of $^{244}$Pu observed across this time window, which is probed by meteoritic analysis and deep ocean crust measurements (*73*), involves rare *r*-process events that would occur every ~500 Myr, on average, near the Sun. This low frequency has been derived independently (*74*), while other investigators have obtained a lower value of ~100 Myr when assuming a site four times more frequent (*75*).

Time Since the Last *r*-Process Event

Calculating the time elapsed between the last *r*-process event and the condensation of the first solids in the early Solar System, which we refer to as $\Delta t_{\mathrm{LE}}$, allows us to calculate backward the early Solar System $^{129}$I / $^{247}$Cm ratio and recover the nucleosynthesis production of the last event (see above). For the *r*-process, this can be done by predicting the isotopic ratios $^{129}$I / $^{127}$I and $^{247}$Cm / $^{235}$U in the parcel of gas from which the Sun formed, just after that last enrichment event, and determine the time required for the value to decay until they reach the early Solar System composition, which sets $\Delta t_{\mathrm{LE}}$ (*6,76*). Calculating $\Delta t_{\mathrm{LE}}$ also provides a consistency check for the nucleosynthesis models presented in Figs. 2 and S2. Plausible models must reproduce both the $^{129}$I / $^{247}$Cm ratio and the $\Delta t_{\mathrm{LE}}$ derived independently from $^{129}$I / $^{127}$I and $^{247}$Cm / $^{235}$U, $^{129}$I and $^{247}$Cm are co-produced by the same event.

Given the rarity of *r*-process events in the solar neighbourhood (see above), we start by considering that $^{129}$I and $^{247}$Cm in the early Solar System mostly came from one *r*-process event only (we quantify the likelihood of this assumption in the next section). In this case, their abundance ratio in the parcel of gas that constituted the pre-solar matter at time $T_{\mathrm{gal}}$ when the event occurred (measured from the birth of the Galaxy) is proportional to the nucleosynthesis yields (Y ) of $^{129}$I ($Y_{129\mathrm{I}}$) and $^{247}$Cm ($Y_{247\mathrm{Cm}}$) synthesized in the event. $T_{\mathrm{gal}}$ ranges between 8 and 9 billion years and was calculated by subtracting the age of the Solar System to the age of the Milky Way, assuming its formation began within one billion years following the Big Bang. The abundance of the stable $^{127}$I includes contributions from all the previous *r*-process events that enriched the parcel of gas. Assuming an average time interval $\langle\delta\rangle$ between *r*-process events, the total number of $^{127}$I isotopes within that parcel of gas is proportional to

$$Y_{^{127}\mathrm{I}}\left(\frac{T_{\mathrm{gal}}}{\langle\delta\rangle}\right). \tag{S1}$$

The iodine isotopic ratio can be written as (*6,76*)

$$\frac{^{129}\mathrm{I}}{^{127}\mathrm{I}} = K_{\mathrm{I}} \left( \frac{Y_{^{129}\mathrm{I}}}{Y_{^{127}\mathrm{I}}} \right) \left( \frac{\langle \delta \rangle}{T_{\mathrm{gal}}} \right), \tag{S2}$$

where $K_I$ is a correction factor extracted from galactic chemical evolution models to account for the temporal evolution of the star formation rate in the Galaxy and the amount of stable isotopes locked inside stellar remnants (*77,78*), as well as for the amount of stable isotopes ejected outside the Galaxy by galactic outflows (*14*).

The abundance of $^{235}$U at $T_{gal}$, because of its long half-life relative to $^{129}$I and $^{247}$Cm, includes the contribution of several *r*-process events, but fewer than the total number that polluted the parcel of gas. Following published methods [(*6*), their equation 12], and considering that the half-life of $^{235}$U is almost an order of magnitude shorter than $T_{gal}$, the abundance of $^{235}$U will evolve toward a steady state instead of piling up like a stable isotope (*16*). Under those circumstances, the abundance of $^{235}$U by the time the Solar System forms is expected to be [(*6*), their equation 13] proportional to

$$Y_{^{235}\mathrm{U}} \left( \frac{1}{1 - e^{-\langle \delta \rangle / \tau_{^{235}\mathrm{U}}}} \right), \tag{S3}$$

which leads to an isotopic ratio of

$$\frac{^{247}\mathrm{Cm}}{^{235}\mathrm{U}} = K_{\mathrm{U}} \left( \frac{Y_{^{247}\mathrm{Cm}}}{Y_{^{235}\mathrm{U}}} \right) \left( 1 - e^{-\langle \delta \rangle / \tau_{^{235}\mathrm{U}}} \right), \tag{S4}$$

where $K_U$ is equivalent to $K_I$ but taking into account that $^{235}$U is radioactive (*14*).

Fig. S3 shows our results for $\Delta t_{LE}$ as a function of the recurrence time $\langle \delta \rangle$ between *r*-process events, using the correction factors $K_I = 2.3$ and $K_U = 1.9$ from a galactic chemical evolution model (*14*), and using the yields of three nucleosynthesis models presented in Fig. S2. The error bars combine three sources of uncertainties: the 2σ uncertainties in the half- life of $^{129}$I and $^{247}$Cm (± 10 Myr), the range of $K_I$ (1.6 − 5.7) and $K_U$ (1.4 − 3.2) which accounts for galactic evolution uncertainties (*14*) (± 15 and 10 Myr, respectively), and the possibility that stable and long-lived *r*-process species are distributed heterogeneously among solar-metallicity stars within a factor of two (*79–81*) (± 15 Myr). Within a plausible range of $\langle \delta \rangle$ (see above), the time between the condensation of solids in the early Solar System and the last *r*-process event ranges between 100 and 200 Myr. This range is consistent with previous estimates (*14,74–76*).

For the dynamical ejecta NS-NS dyn. (B) and the disk ejecta NS-NS disk 1, the times derived from $^{129}$I / $^{127}$I and $^{247}$Cm / $^{235}$U are more consistent with each other than for the MR SN model. Using the different nuclear physics input shown in Fig. S2 does not alter those conclusions. Fig. S3C shows that the physical conditions in NS-NS disk 1 can synthesize $^{129}$I and $^{247}$Cm in such a way that both isotopes trace back the same *r*-process event, in addition to being consistent with the $^{129}$I / $^{247}$Cm meteoritic ratio.

This calculation assumes that all *r*-process events generate the same nucleosynthesis product every time. Uncertainties in the nuclear physics and in the physical conditions of *r*-process calculations are too large (cf. NS-NS disks 1, 2, and 3 in Fig. S2) to examine yield variations. Our calculations do not consider explicitly any transport of material in the interstellar medium, but assume that all the isotopes are diluted by the same factor. Because such mixing processes are complex (*82–85*) and cannot be captured with non-hydrodynamic approaches, we regard the timescales estimated above as first approximations.

Uncertainty on the Meteoritic $^{129}I$ / $^{247}Cm$ Ratio

The 42% uncertainty on early Solar System $^{129}I$ / $^{247}Cm$ ratio (Table 1) is dominated by the 40% (2σ) uncertainty on the Solar System elemental abundance of I (*5*). The $^{247}Cm$ / $^{235}U$ ratio, used in the derivation of $^{129}I$ / $^{247}Cm$, is determined by measurements of one calcium-aluminium-rich inclusion (CAI) only. The 2σ uncertainty on the elemental U abundance is 16% (*5*), and the uncertainty on the determination of the isotopic fraction (i.e., of $^{235}U$) is negligible in comparison (*5*). Because the homogeneity of $^{129}I$ / $^{127}I$ in the early Solar System is well established (*7*), we assume that $^{247}Cm$ / $^{235}U$ was also homogeneously distributed.

To compare with the nucleosynthesis predictions, the $^{129}I$ / $^{247}Cm$ ratio derived from meteoritic analysis must reflect the nucleosynthesis production of the last *r*-process event. As described in the main text, because $^{129}I$ and $^{247}Cm$ have very similar half lives, their ratio should not vary between the *r*-process event that produced them and the formation of the first solids in the early Solar System. This time interval can be determined using the $^{129}I$ / $^{127}I$ and $^{247}Cm$ / $^{235}U$ ratios, and is estimated to be between 100 and 200 million years (Myr) (see above).

The half-lives of $^{129}I$ and $^{247}Cm$ have 2σ uncertainties of 5% and 6% *(11–12)*, respectively, and their isotopic ratio changes with an equivalent half-life

$$\tau_{\mathrm{eq}} = \frac{\tau_1 \tau_2}{(\tau_2 - \tau_1)}, \quad \text{(S5)}$$

where $\tau_1$ and $\tau_2$ are the half-lives of $^{129}I$ and $^{247}Cm$, respectively. We derived the probability distribution function of $\tau_{\mathrm{eq}}$ using Monte Carlo calculations that randomly sampled the half- lives of $^{129}I$ and $^{247}Cm$ assuming a normal distribution for each, based on their 2σ uncertainty (see Table 1). We found that $\tau_{\mathrm{eq}}$ ranges from ~100 Myr to several billion years (Fig. S4). The short timescales covered by the tail of the distribution implies the possibility that the ejected $^{129}I$ / $^{247}Cm$ ratio could have deviated from its original value before being locked into solids.

To account for this possible variation, we calculated the Solar System value backward in time for 200 Myr using a Monte Carlo approach. In each run, we randomly sampled the equivalent half-life from the probability distribution of $\tau_{\mathrm{eq}}$, as well as the early Solar System value from a normal distribution based on the 2σ uncertainty presented in Table 1. After running this calculation 10 million times, we found values for the $^{129}I$ / $^{247}Cm$ ratio between 264 and 616 at 1σ, and between 166 and 922 at 2σ (Fig. S5). These confidence intervals are shown in Fig. 2 and represent the uncertainty in the $^{129}I$ / $^{247}Cm$ ratio after the last *r*-process event prior the formation of the Solar System. We regard these uncertainties as conservative, because our calculations suggest that the last *r*-process event likely occurred less than 200 Myr before the condensation of the first solids in the early Solar System [(*16*) and Fig. S3].

The uncertainty on $^{129}$I / $^{247}$Cm is only weakly affected by the uncertainties on the time since the last *r*-process event. Assuming 100 Myr instead of 200 Myr in our Monte Carlo calculation changes the uncertainty by 40%, while the values of $^{129}$I / $^{127}$I and $^{247}$Cm / $^{235}$U change by two orders of magnitude.

## Supplementary Text

Deriving Physical Conditions from the $^{129}$I / $^{247}$Cm Ratio

As shown in Fig. S2, the isotopic ratio of the dynamical ejecta NS-NS dyn. (B) (*20*) is lower than the meteoritic value by roughly an order of magnitude. However, this model contains several individual trajectories (i.e., sets of physical conditions, see above) that cover a wide range of ratios, both lower and higher than the meteoritic value. Considering all those trajectories as a collection of possible physical conditions from which we can sample, we searched for a combination of trajectories that would lead to the early Solar System value of 438 ± 92 (within 1σ uncertainty), as reported in Table 1. Although all trajectories come from a NS-NS merger simulation, the goal of this exercise is to investigate what distribution of physical conditions can generate the meteoritic ratio. We chose the NS-NS dyn. (B) simulation for this purpose because of the wider range and larger number of available trajectories, compared to the other simulations shown in Fig. S2.

We employed an iterative random sampling procedure (*86*). In each iteration, we randomly selected ten trajectories, summed the abundances of $^{129}$I and $^{247}$Cm, and calculated the integrated $^{129}$I / $^{247}$Cm abundance ratio. If the ratio was in the allowed range of 438 ± 92, the ten trajectories were added to the sampled ejecta, and this process was repeated until the integrated $^{129}$I / $^{247}$Cm ratio of the sampled ejecta did not change by more than a factor of $10^{-6}$ when ten new trajectories were added. This sampled ejecta then represents a subset of trajectories that, once combined together, fits the early Solar System ratio.

In Fig. S1, we show the distribution of electron fractions $Y_e$ of the sampled ejecta. This quantity probes the amount of available neutrons, defined as,

$$Y_e = \frac{n_p}{n_n + n_p}, \quad \text{(S6)}$$

where $n_p$ and $n_n$ are the particle density of protons and neutrons, respectively. The lower $Y_e$ is, the more neutron-rich are the conditions. As shown in Fig. S1, the $Y_e$ distribution of the sampled ejecta is less neutron-rich than the original dynamical ejecta. The bulk of the fitted $Y_e$ distribution is in between 0.2 and 0.3, with a very small contribution on the very neutron-rich side ($Y_e <$ 0.15), which is the opposite of the original distribution from which we sampled. These sampled ejecta are not associated with any *r*-process site in particular, they only represent an example of what conditions can reproduce the $^{129}$I / $^{247}$Cm ratio in the early Solar System. Nevertheless, when we compare to the ejecta of NS-NS disk 1 (*21*), which is consistent with meteoritic $^{129}$I / $^{247}$Cm (see Fig. S2), we find the $Y_e$ distribution to peak at a value that is similar to what is given by our sampling procedure.

Meteoritic Iodine Abundance Measurements

In this work, we adopt Solar System compositions that are typically used in astronomy (*5*) to derive the $^{129}$I / $^{247}$Cm abundance ratio. Alternative halogen measurements (chlorine, bromine, and iodine) (*17*) suggest abundances that are systematically lower than what is adopted in our work (*5*). The abundances for the early Solar System are derived from CI chondrites. In these meteorites, the spatial distribution of halogens such as iodine is highly heterogeneous (*87,88*). The typical threshold mass required for a meteorite sample to ensure representative solar abundance measurements is ~1g (*89*) or more for halogens. The sample used in alternative halogen measurements for chlorine, bromine, and iodine (*17*) was less than ~3mg, which makes it possible to miss relevant halogen dense regions. In addition, the size of the grains hosting halogens in meteorites can have an impact on measurements when using the neutron irradiation technique adopted for the alternative measurements. The noble gases produced from halogens via irradiation, which are used to derive halogen abundances, are highly volatile and can be lost from the system by diffusion. In ordinary chondrites, the average size of halogen bearing minerals is ~200μm (*89*), which is large enough to minimize the importance of this effect. For these ordinary chondrites, the alternative measurements reproduced the range of halogen abundances from previous studies. However, for carbonaceous chondrites and for the single CI chondrite sample used in the alternative study (*17*) for early Solar System considerations, the average grain size is less than 1μm (*87*), which make halogen measurements on these meteorites more affected by the possible losses described above.

Independent of meteorite analysis, the solar abundance of chlorine can be inferred from spectroscopic observations of the Sun's surface (*90,91*). These inferred abundances are consistent with the composition of nearby stars (*92,93*) and with the meteoritic chlorine abundance commonly adopted in astronomy (*5*,*94*), but are higher than the alternative measurements (*17*). Considering that the abundances of all three halogens reported by these alternative measurements are systematically lower than previous studies, if their chlorine abundance was underestimated, their iodine abundance would also be underestimated. The alternative study derived relative halogen abundances (Br/Cl and I/Cl ratios) in primitive meteorites that are consistent with previous studies and found that the Earth's mantle and crust have similar halogen ratios to chondrites (*17*). If the absolute chlorine abundance was scaled up to match the abundance observed for the Sun, the iodine abundance would also need to be scaled up according to the above halogen ratios. By doing so, the alternative iodine value would increase from $0.118 \pm 0.015$ (normalized to $\mathrm{Si} = 10^6$) to $0.72 \pm 0.43$, which overlaps with the $1.1 \pm 0.22$ range adopted in our study (*5*).

An additional point to consider is the overall abundance pattern inferred from meteoritic measurements. Adopting the alternative lower halogen abundances generates significant dips in the Solar System abundance pattern that are not commonly seen with nucleosynthesis calculations. For bromine, we did not find galactic chemical evolution simulations that predict an underproduction of bromine relative to its neighbour elements selenium and krypton (*95–98*). For the *r* process, when isotopes lighter and heavier than iodine are both significantly produced in our calculations, the abundance patterns typically predict iodine to smoothly fall between neighbouring species, as is the case with the abundances adopted in our study (*5*). But in the very neutron-rich conditions considered in Tables S2 and S3 with PRISM, we found a few trajectories for which the abundance pattern starts at a mass number slightly above A = 127 and extends up to the actinides. Additionally, trajectories for which the synthesis can reach up to a mass number slightly below A = 127 can be found in less neutron-rich conditions. It is a mixture of these two types of ejecta that could potentially generate a dip in iodine. The mentioned trajectories found in our calculations, however, represent a small portion of the total ejecta. Therefore although we were able to identify a path toward generating the dip at A = 127, it would require a distribution for the total ejecta which is not currently common in nucleosynthesis calculations.

Probability of Proving One *r*-Process Event

Metal-poor stars exhibit different levels of actinide enrichment relative to rare-Earth peak elements, as shown by their elemental ratio between thorium and europium (*86,99,100*). This indirectly suggests that actinide isotopes such as $^{247}$Cm are not always synthesized in the same amount from one *r*-process event to another. Verifying quantitatively that the early Solar System $^{129}$I / $^{247}$Cm ratio most likely encodes the signature of one event is necessary to connect this ratio to *r*-process nucleosynthesis calculations.

The stochastic evolution of radioactive isotopes in a given parcel of gas within the interstellar medium has previously been investigated (*16*) as a function of their half-life $\tau$ and the average time interval $\langle\delta\rangle$ between two consecutive enrichment events. They found that the probability of probing one event with a radioactive isotope varies from 0 to 100% when the timescale ratio $\tau$ / $\langle\delta\rangle$ varies from 1 to 0.01. Because the half-life of $^{129}$I and $^{247}$Cm is 15.6 Myr, and because $\langle\delta\rangle$ for *r*-process events is probably between 100 and 500 Myr (see above), the $\tau$ / $\langle\delta\rangle$ ratio is roughly between 0.05 and 0.2. This range implies that for more than 90% of the time [(*16*), their figure 10], 90% of the abundances of $^{129}$I and $^{247}$Cm in the interstellar medium were produced by only one event.

A single *r*-process event could account for 50 to 100% of all $^{247}$Cm present in the early Solar System [(*75,101*), who use a similar approach to (*15*)]. This last *r*-process event could have contributed to about 80% the amount of $^{244}$Pu present in the early Solar System, leaving a 20% contribution from other previous events (*74*). Because $^{244}$Pu has a half-life five times longer than $^{129}$I and $^{247}$Cm, we expect the latter isotopes to be even more dominated by the last *r*-process event.

Possible Contribution by Two *r*-Process Events

To test our conclusion that $^{129}$I / $^{247}$Cm only probes one event, and the implications of probing multiple events, we performed a Monte Carlo calculation of the possible contributions from the two last *r*-process events that preceded the formation of the Solar System, using a subset of the nucleosynthesis calculations presented in Fig. S2. For these tests, we used the dynamical ejecta NS-NS dyn. (B), the MR-SN , and the disk NS-NS disk 1, as these three models predict very different $^{129}$I / $^{247}$Cm ratios from each another (see Fig. S2). This allows us to quantify the probability that a mixture of two events could match the observed ratio, even if the single events do not. For these models, we considered the ABLA07 fission fragment distribution model and the DZ10 mass model to ensure that NS-NS disk 1 reproduces the meteoritic ratio. This allows us to quantify the chance of mixing a single non-matching event with a matching event, and remaining within the uncertainties of the observed ratio. We assumed that all events are independent and can occur at different times, even if disk and dynamical ejecta are ejected together in the case of compact binary mergers. To consistently mix the ejecta of the two *r*-process events we account for the total mass of $^{129}$I and $^{247}$Cm ejected by each event (see Table S5).

We start by randomly sampling two *r*-process events and separating them by a time delay $\delta$. This delay is randomly sampled from probability distribution functions (*16*), which depend on the nature and average frequency of the enrichment source. We decay the ejecta of the first event for a time $\delta$ using an equivalent half-life $\tau_{eq}$ drawn randomly from the distribution derived above, based on the half-life uncertainties of $^{129}$I and $^{247}$Cm. Then, we mix the decayed ejecta of the first event with the ejecta of the second event (i.e., the last event before the formation of the Solar System). We compare the mixture against the meteoritic ratio, which was already corrected backward in time to the last *r*-process event (see above). We account for the probability distribution function of the meteoritic value, using a Monte Carlo approach. This means the scenario has more chance to be classified as successful if its $^{129}$I / $^{247}$Cm abundance ratio is close to the center of the meteoritic distribution, while it has less chance if its ratio falls on the tail of the meteoritic distribution (e.g., on the edge of the 2σ uncertainty). We repeated this process a million times to derive the probability of each successful scenario.

The results of all successful scenarios are summarized in Table S6. Two percent estimates are reported. The first were calculated by minimizing the chance of pollution by two events using an average frequency of 1 event per 500 Myr and a meteoritic ratio decayed backward for 100 Myr. The second were calculated by maximizing the chance of pollution by two events using a higher frequency of 1 event per 100 Myr. A higher event frequency enhances the chance of generating two events close in time.

We draw two conclusions from this calculation. First, there is less than 6% chance that two non-disk events combine their ejecta to generate the observed value. Second, if the ejecta of one disk is mixed with another type of ejecta, the probability for the disk to contribute to more than 99% the amount of $^{129}$I and $^{247}$Cm in the early Solar System is between ~51 and 95%. Those probabilities increase to ~84 and 98% for contributions of > 90%. The overall probability of probing two events with production ratios different from the meteoritic value is low.

We also consider that the full ejecta of a single compact binary merger is composed of both dynamical and disk ejecta (*3,102*). NS-NS disk 1 alone is consistent with the meteoritic value, and the uncertainties in the measured and modeled values allow a mixture of dynamical and disk ejecta within the uncertainties. Although this calculation is affected by nuclear physics uncertainties, we select as a test case the set of nuclear physics input that produces the highest ratio for both NS-NS disk 1 and the dynamical ejecta NS-NS dyn. (B). We found that the maximum allowed contribution of dynamical ejecta (to generate a ratio at the 2σ lower limit of the meteoritic value) is 46%, in mass fraction. NS-NS merger simulations typically find that dynamical ejecta contribute less than 50% to the total mass ejected (*1,103*). Additionally, dynamical ejecta alone typically overproduce the Th/Eu elemental ratio derived from the surface of old stars, which suggests that very neutron-rich conditions cannot entirely be responsible for the enrichment of these stars (*86*).

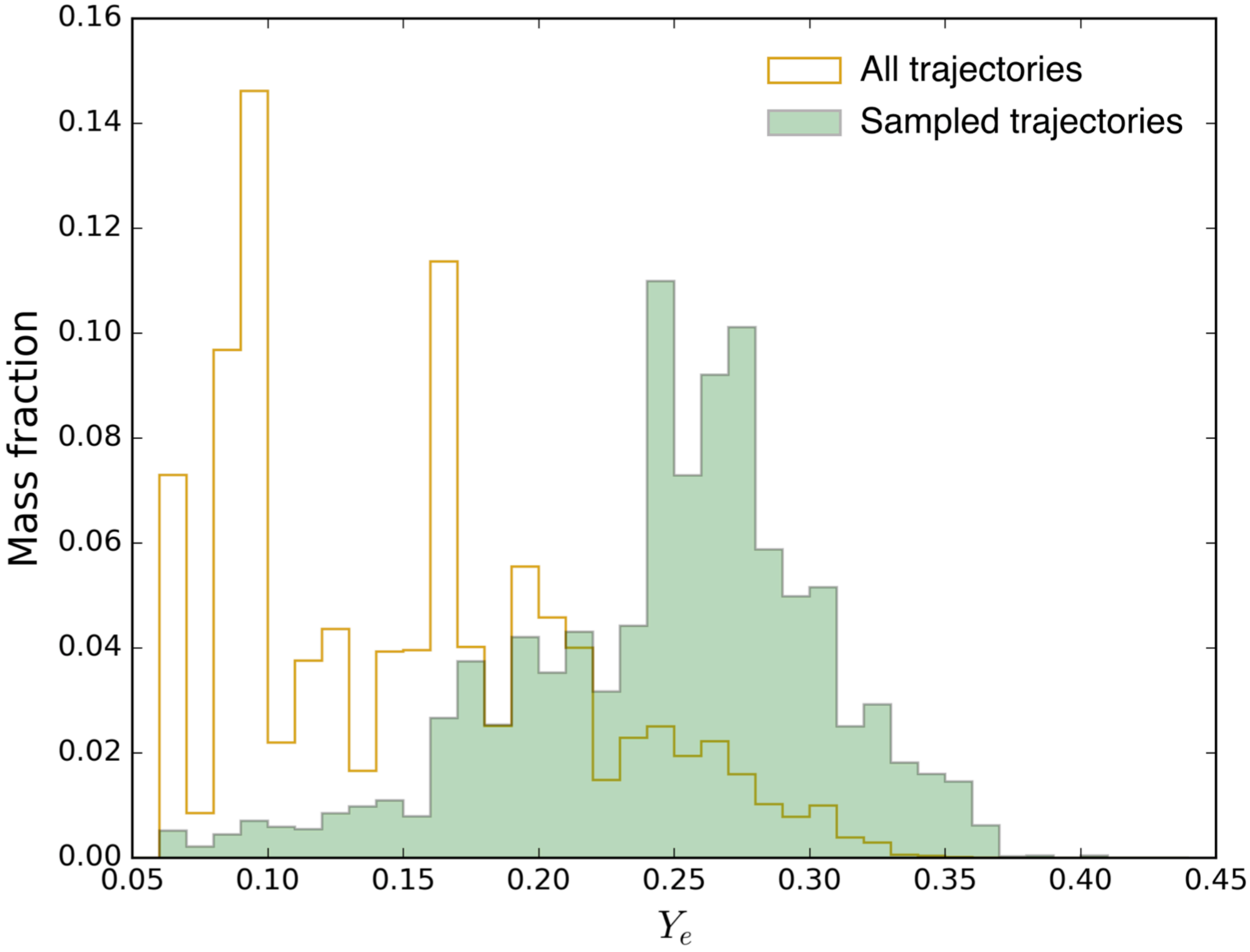

**Fig. S1. Example of the electron fraction ($Y_e$) distribution within $r$-process ejecta.** The ejecta corresponds to the dynamical ejecta of NS-NS dyn. (B) (*20*). The yellow outline histogram shows the original $Y_e$ distribution of the total ejecta, consisting of all trajectories taken at a temperature of 8 GK. The green histogram shows the distribution of the sum of a subset of trajectories that successfully reproduced the early Solar System meteoritic $^{129}$I/$^{247}$Cm ratio of $438 \pm 92$. See Supplementary Text for details.

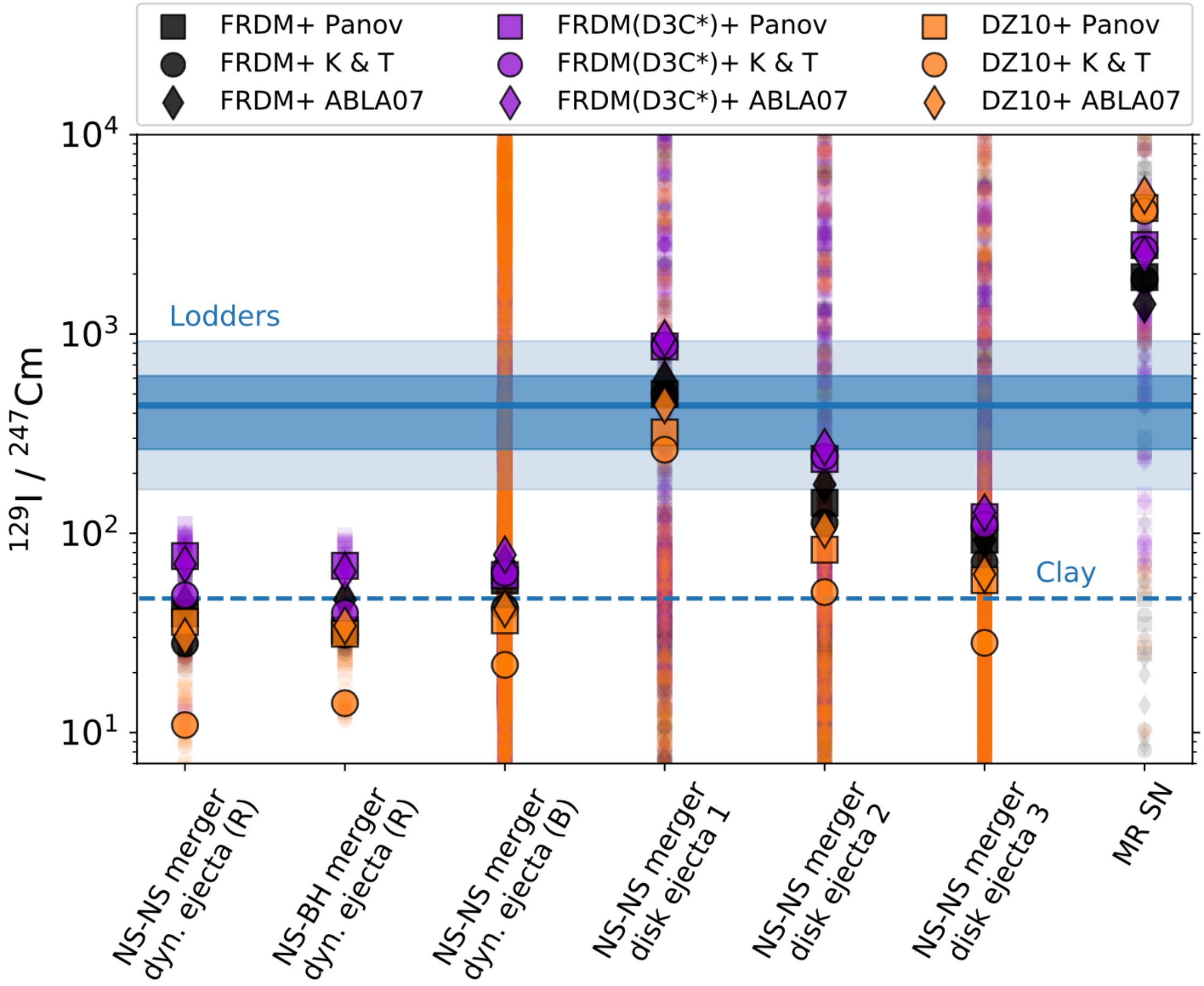

**Fig. S2.** $^{129}$**I** / $^{247}$**Cm abundance ratios predicted by theoretical** $r$**-process models.** This figure is a more detailed version of Fig. 2, where the different symbol shapes and colors denote different nuclear physics input (see Methods). Ratios from the individual trajectories are shown as small, transparent symbols. The total ejecta are represented by the larger, opaque symbols.

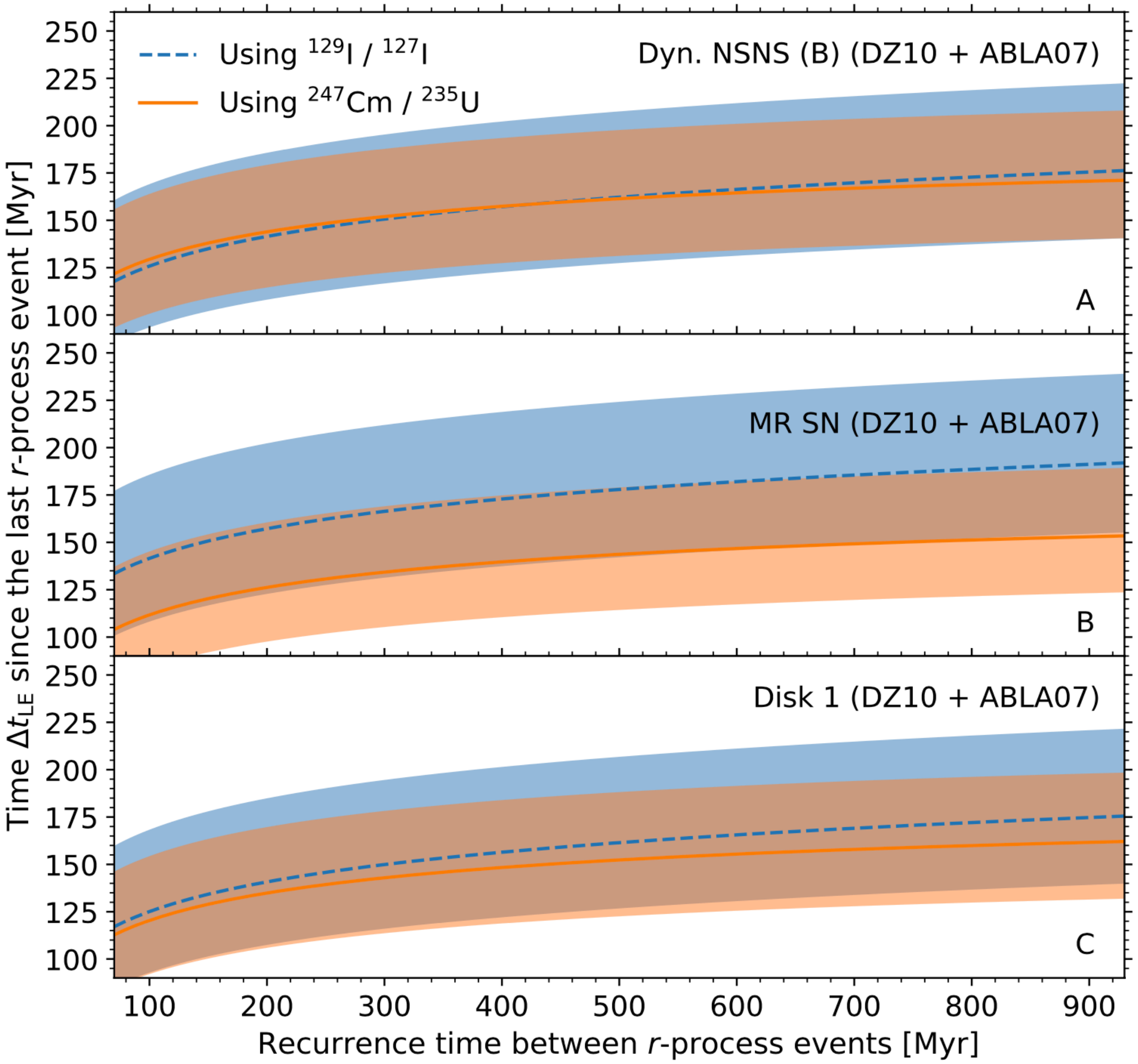

**Fig. S3. Time $\Delta t_{\mathrm{LE}}$ elapsed between the last $r$-process event and the formation of the Solar System.** Each panel used a different nucleosynthesis model to generate the last $r$-process enrichment of the pre-solar nebula. Panel A: Dyn. NSNS (B). Panel B: MR SN. Panel C: NSNS disk 1. The dashed blue and solid orange lines show our calculations using the abundances of $^{129}$I and $^{247}$Cm, respectively, which are independent tracers. The shaded bands surrounding each line indicate the uncertainty, including the $2\sigma$ uncertainty due to uncertainties in the half-life of $^{129}$I and $^{247}$Cm, the uncertainties in the galactic chemical evolution models, and the inhomogeneities of $r$-process elements in the interstellar medium near solar metallicity. See Supplementary Text for details.

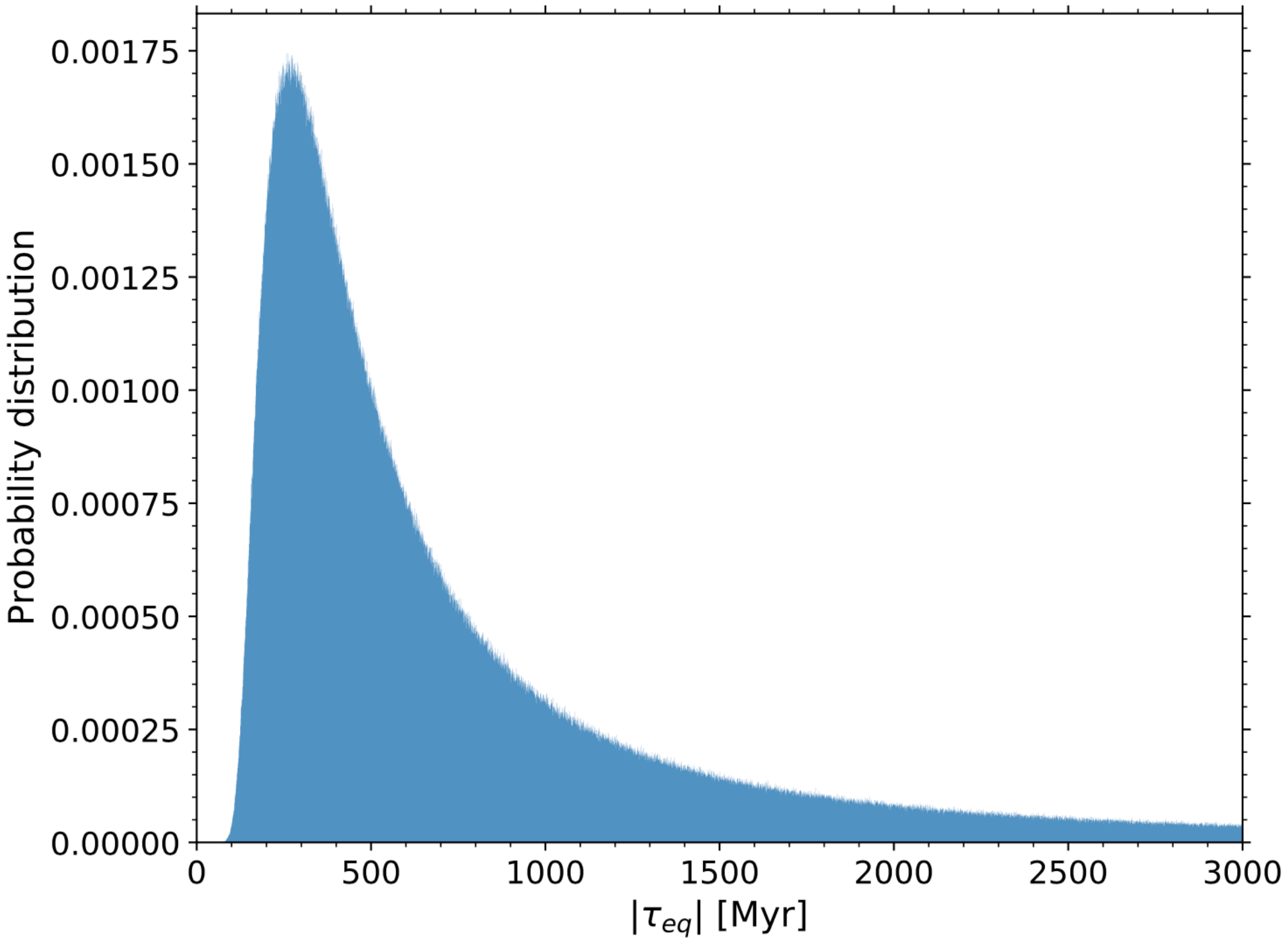

**Fig. S4. Probability distribution of the equivalent half-life ($\tau_{eq}$) of $^{129}$I / $^{247}$Cm**. The range of values reflects the uncertainties on the half-lives of $^{129}$I and $^{247}$Cm (Table 1).

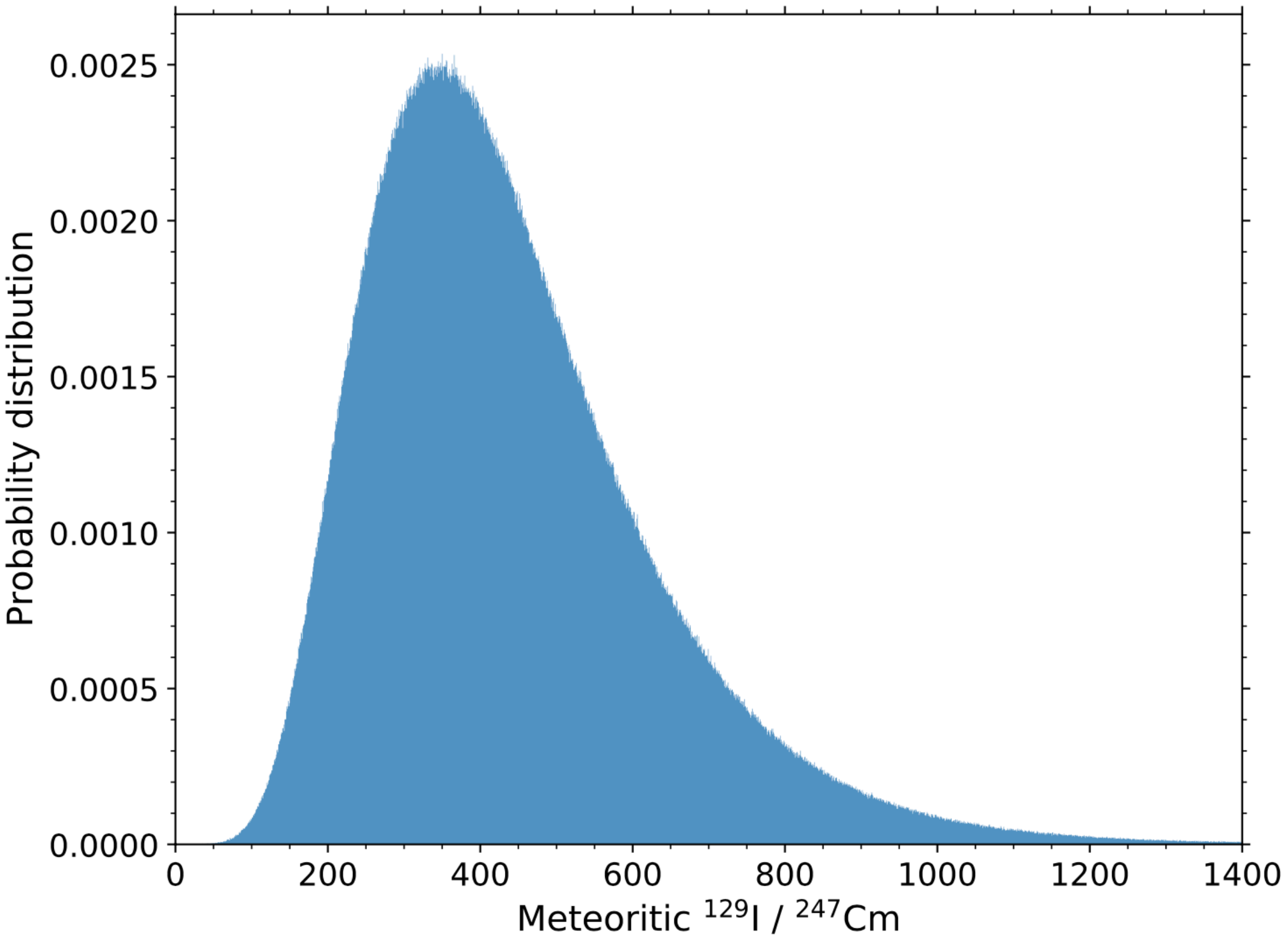

**Fig. S5. Probability distribution of the meteoritic $^{129}$I / $^{247}$Cm abundance ratio**. The distribution represents the pre-solar nebula 200 Myr before the formation of the Solar System, and the range of values reflects the uncertainties on the Solar System abundances ratios and the half-lives of $^{129}$I and $^{247}$Cm.

**Table S1. Details of the hydrodynamical models of $r$-process scenarios shown in Figs. 2 and S2.** The labels in the first column correspond to those shown in the figures. The setup column contains information on the the initial parameters of the simulations: masses and equation of state (EOS) for binary mergers; progenitor mass and initial magnetic field strength $B_0$ for the magneto-rotational supernova; torus and BH mass, initial entropy $s_0$ in units of Boltzmann constant $k_B$ per baryon, and viscosity parameter $\alpha$ for the disk scenarios. All masses are expressed in units of solar mass [$M_\odot$]. The EOSs Shen and SFHO are named after the authors that published the work.

| Label | Type of scenario | Setup | Reference | Name in reference |
|---|---|---|---|---|
| NS-NS (R) | NS-NS merger dynamical ejecta | 1.0 $M_\odot$ + 1.0 $M_\odot$, Shen EOS | (*19,46,104*) | Run 1 |
| NS-BH (R) | NS-BH merger dynamical ejecta | 1.4 $M_\odot$ + 5.0 $M_\odot$, Shen EOS | (*19,46,104*) | Run 22 |
| NS-NS (B) | NS-NS merger dynamical ejecta | 1.25 $M_\odot$ + 1.25 $M_\odot$, SFHO EOS | (*20*) | SFHO-M1.25 |
| MR SN | Magneto-rotational supernova | 15 $M_\odot$, $B_0 = 5 \times 10^{12}$ G | (*18*) | w/ $\nu$ heating |
| Disk 1 | NS-NS merger disk ejecta | 0.03 $M_\odot$ + 3.0 $M_\odot$, $s_0 = 8\ k_B$/b, $\alpha = 0.03$ | (*21*) | S-def |
| Disk 2 | NS-NS merger disk ejecta | 0.03 $M_\odot$ + 3.0 $M_\odot$, $s_0 = 6\ k_B$/b, $\alpha = 0.03$ | (*21*) | s6 |
| Disk 3 | NS-NS merger disk ejecta | 0.03 $M_\odot$ + 3.0 $M_\odot$, $s_0 = 8\ k_B$/b, $\alpha = 0.10$ | (*21*) | $\alpha$0.10 |

**Table S2. The $^{129}$I/$^{247}$Cm ratio calculated with PRISM using different mass models (see Table S4).** All calculations are based on NS-NS merger dynamical ejecta simulations (*19,46,104*). When labeled (D3C*, which stands for adjusted model with Density-Dependent and Derivative Couplings), we replaced our default $\beta$-decay rates (*67*) by alternatives (*42*). The numbers in the first column indicate the masses of the neutron stars in units of solar masses. The third column shows the range covered by all trajectories, while the fourth column shows the mass weighted average of those trajectories.

| NS-NS merger scenario | Mass model | $^{129}$I/$^{247}$Cm Range | $^{129}$I/$^{247}$Cm Weighted average |
|---|---|---|---|
| NS1.2 - NS1.4 | TF (D3C*) | 113.82 – 615.01 | 360.69 |
| | ETFSI-Q | 42.21 – 127.57 | 92.29 |
| | ETFSI | 37.94 – 108.56 | 80.04 |
| | FRDM2012 (D3C*) | 9.01 – 98.78 | 67.19 |
| | ETFSI (D3C*) | 3.55 – 126.40 | 54.01 |
| | SLy4 | 8.79 – 37.31 | 28.62 |
| | TF | 11.77 – 25.74 | 21.79 |
| | HFB-17 (D3C*) | 2.34 – 36.20 | 16.20 |
| | FRDM2012 | 5.10 – 23.69 | 13.72 |
| | UNEDF0 | 8.28 – 17.50 | 13.72 |
| | HFB-21 | 4.99 – 16.05 | 9.24 |
| | HFB-17 | 6.18 – 13.21 | 9.11 |
| | DZ33 | 5.16 – 8.74 | 6.43 |
| | FRDM1995 | 2.44 – 8.31 | 5.52 |
| NS1.4 - NS1.4 | TF (D3C*) | 223.93 – 599.49 | 387.29 |
| | ETFSI | 58.71 – 86.03 | 69.06 |
| | ETFSI-Q | 53.62 – 83.21 | 67.60 |
| | FRDM2012 (D3C*) | 15.72 – 89.16 | 50.14 |
| | ETFSI (D3C*) | 7.55 – 109.17 | 38.91 |
| | SLy4 | 15.37 – 33.35 | 27.18 |
| | TF | 15.22 – 23.36 | 19.83 |
| | FRDM2012 | 11.39 – 28.60 | 16.65 |
| | HFB-21 | 6.82 – 15.45 | 12.18 |
| | UNEDF0 | 11.19 – 13.06 | 11.97 |
| | HFB-17 (D3C*) | 2.66 – 30.54 | 10.83 |
| | HFB-17 | 8.45 – 10.73 | 9.33 |
| | DZ33 | 6.09 – 8.58 | 7.40 |
| | FRDM1995 | 2.46 – 7.48 | 4.33 |

**Table S3. Same as Table S2, but for NS-BH mergers.**

| NS-BH merger scenario | Mass model | $^{129}$I/$^{247}$Cm Range | $^{129}$I/$^{247}$Cm Weighted average |
|---|---|---|---|
| NS1.4 - BH5 | TF (D3C*) | 123.91 – 231.83 | 187.56 |
| | ETFSI | 43.92 – 282.60 | 83.01 |
| | ETFSI-Q | 41.04 – 321.49 | 81.59 |
| | SLy4 | 10.07 – 22.15 | 14.94 |
| | TF | 12.15 – 15.61 | 13.80 |
| | FRDM2012 (D3C*) | 8.65 – 15.61 | 13.16 |
| | UNEDF0 | 8.43 – 14.01 | 10.92 |
| | HFB-21 | 5.30 – 14.05 | 10.43 |
| | HFB-17 | 7.39 – 22.95 | 10.19 |
| | FRDM2012 | 5.37 – 13.43 | 9.27 |
| | DZ33 | 5.20 – 13.11 | 7.05 |
| | ETFSI (D3C*) | 3.70 – 8.71 | 6.02 |
| | HFB-17 (D3C*) | 2.52 – 5.04 | 3.45 |
| | FRDM1995 | 2.31 – 17.19 | 3.42 |
| NS1.4 - BH10 | TF (D3C*) | 143.51 – 531.01 | 235.37 |
| | ETFSI-Q | 39.80 – 567.25 | 84.49 |
| | ETFSI | 38.39 – 73.23 | 61.19 |
| | FRDM2012 (D3C*) | 9.15 – 88.10 | 23.80 |
| | SLy4 | 12.37 – 105.77 | 23.33 |
| | ETFSI (D3C*) | 3.96 – 116.48 | 17.82 |
| | TF | 11.51 – 21.94 | 15.48 |
| | FRDM2012 | 7.92 – 17.66 | 12.17 |
| | HFB-21 | 6.49 – 15.33 | 11.55 |
| | UNEDF0 | 6.52 – 15.29 | 11.27 |
| | HFB-17 | 8.00 – 14.37 | 9.75 |
| | DZ33 | 4.63 – 13.39 | 7.44 |
| | HFB-17 (D3C*) | 2.47 – 32.40 | 6.38 |
| | FRDM1995 | 2.47 – 36.30 | 4.97 |

**Table S4. Mass models adopted for the nucleosynthesis calculations with PRISM.**

| Mass model | Acronym definition | Reference |
|---|---|---|
| FRDM2012 | Finite Range Droplet Model | (*52*) |
| FRDM1995 | Finite Range Droplet Model | (*39*) |
| DZ33 | Duflo & Zuker | (*43*) |
| TF | Thomas-Fermi | (*53*) |
| ETFSI | Extended Thomas-Fermi plus Strutinsky Integral | (*54*) |
| ETFSI-Q | ETFSI with shell quenching | (*55*) |
| HFB-17 | Hartree-Fock-Bogoliubov | (*56*) |
| HFB-21 | Hartree-Fock-Bogoliubov | (*57*) |
| SLy4 | Hartree-Fock with Skyrme force SLy4 | (*58*) |
| UNEDF0 | Universal Nuclear Energy Density Functional | (*59*) |

**Table S5. Mass fraction of $^{129}$I and $^{247}$Cm in the ejecta of the nucleosynthesis models used in Table S6.** All models used ABLA07 for the fission fragment distribution and DZ10 for the mass model. The last column shows the total mass ejected, in units of solar masses [$M_\odot$], which includes all stable and radioactive isotopes.

| Model | Mass fraction $^{129}$I | Mass fraction $^{247}$Cm | Total mass ejected [$M_\odot$] |
|---|---|---|---|
| NS-NS dyn. (B) | $3.25 \times 10^{-2}$ | $1.49 \times 10^{-3}$ | $5.50 \times 10^{-4}$ |
| NS-NS disk 1 | $3.77 \times 10^{-2}$ | $1.64 \times 10^{-4}$ | $1.70 \times 10^{-3}$ |
| MR SN | $6.10 \times 10^{-2}$ | $2.35 \times 10^{-5}$ | $6.72 \times 10^{-3}$ |

**Table S6. Probability for two $r$-process events to combine their ejecta and generate a mixture consistent with the meteoritic $^{129}$I/$^{247}$Cm ratio.** In each cell, the two percentages correspond to the minimum and maximum cases described in the Supplementary Text. The label in the first column of each row, Two disks, One disk, and No disk, refers to the number of disk event(s) among the two $r$-process events. The third and fourth columns show the probability that the disk event contributed more than 90% and 99% of both $^{129}$I and $^{247}$Cm in the early Solar System. Such probabilities are not meaningful for the Two disks scenario because in our calculation the nucleosynthesis products were the same for both disks. For the No disk scenario, the two events individually did not have an $^{129}$I/$^{247}$Cm ratio consistent with the meteoritic ratio.

| Scenario | Occurrence probability | One event contributing more than 90% | One event contributing more than 99% |
|---|---|---|---|
| Two disks | $33-40\%$ | – | – |
| One disk | $66-54\%$ | $98-84\%$ | $95-51\%$ |
| No disk | $1-6\%$ | – | – |